\documentclass[prl,aps,twocolumn,superscriptaddress]{revtex4-1}
\usepackage{graphicx,color}
\usepackage{amsthm}
\usepackage{amsfonts}
\usepackage{algorithmic}
\usepackage{enumerate}
\usepackage{latexsym}
\usepackage{amsmath}
\usepackage{amssymb}
\usepackage[colorlinks=true,citecolor=blue,linkcolor=blue]{hyperref}

\newcommand{\be}{\begin{equation}}
\newcommand{\ee}{\end{equation}}
\newcommand{\ba}{\begin{eqnarray}}
\newcommand{\ea}{\end{eqnarray}}

\newcommand{\COMMENTED}[1]{}

\newcommand{\ob}[1]{{\langle #1\rangle}}

\newcommand{\bfa}{\mathbf{a}}
\newcommand{\bfb}{\mathbf{b}}

\newcommand{\bfi}{\mathbf{i}}

\newcommand{\bfr}{\mathbf{r}}

\emergencystretch=\maxdimen
\begin{document}

\title{Metal-Insulator transition in strained Graphene: A quantum Monte carlo study}

\author{Lufeng Zhang}
\email{lfzhang@bupt.edu.cn}
\affiliation{School of Science, Beijing University of Posts and Telecommunications,
Beijing 100876, China}

\author{Chi Ma}
\affiliation{Department of Physics, Beijing Normal University, Beijing
100875, China}
\author{Tianxing Ma}
\email{txma@bnu.edu.cn}
\affiliation{Department of Physics, Beijing Normal University, Beijing
100875, China}

% Abstract should be written in the present tense and impersonal style (i.e., avoid we), and be at most 200 words long
\begin{abstract}
Motivated by the possibility of a strain tuning effect on electronic properties of graphene, the semimetal-Mott insulator transition process on the uniaxial honeycomb lattice is numerically studied using Determinant Quantum Monte Carlo. As our simulations are based on the half-filled repulsive Hubbard model, the system is sign problem free. Herein, the temperature-dependent DC conductivity is used to characterize electronic transport properties. The data suggest that metallic is suppressed in the presence of strain. More interestingly, within the finite-size scaling study, a novel antiferromagnetic phase arises at around $U\sim U_{c}$. Therefore, a phase diagram generated by the competition between interactions and strain is established, which may help to expand the application of strain effect on graphene.
\end{abstract}

\maketitle
% Text: Please use section headings and subheadings as specified below. For communications, all section headings apart from Experimental Section should be removed
% Please make the first reference to a display item bold: \textbf{Figure 1}
% Do not abbreviate Figure, Equation, etc.; display items are always singular, i.e., Figure 1 and 2.
% Equations are always singular, i.e., Equation 1 and 2, and should be inserted using the {equation} environment, not as graphics
% Please do not use footnotes in the text, additional information can be added to the Reference list.

\section{Introduction}
\begin{figure}
\center{%\includegraphics[width=\linewidth]{Fig1_structure}
\includegraphics[width=3.0in]{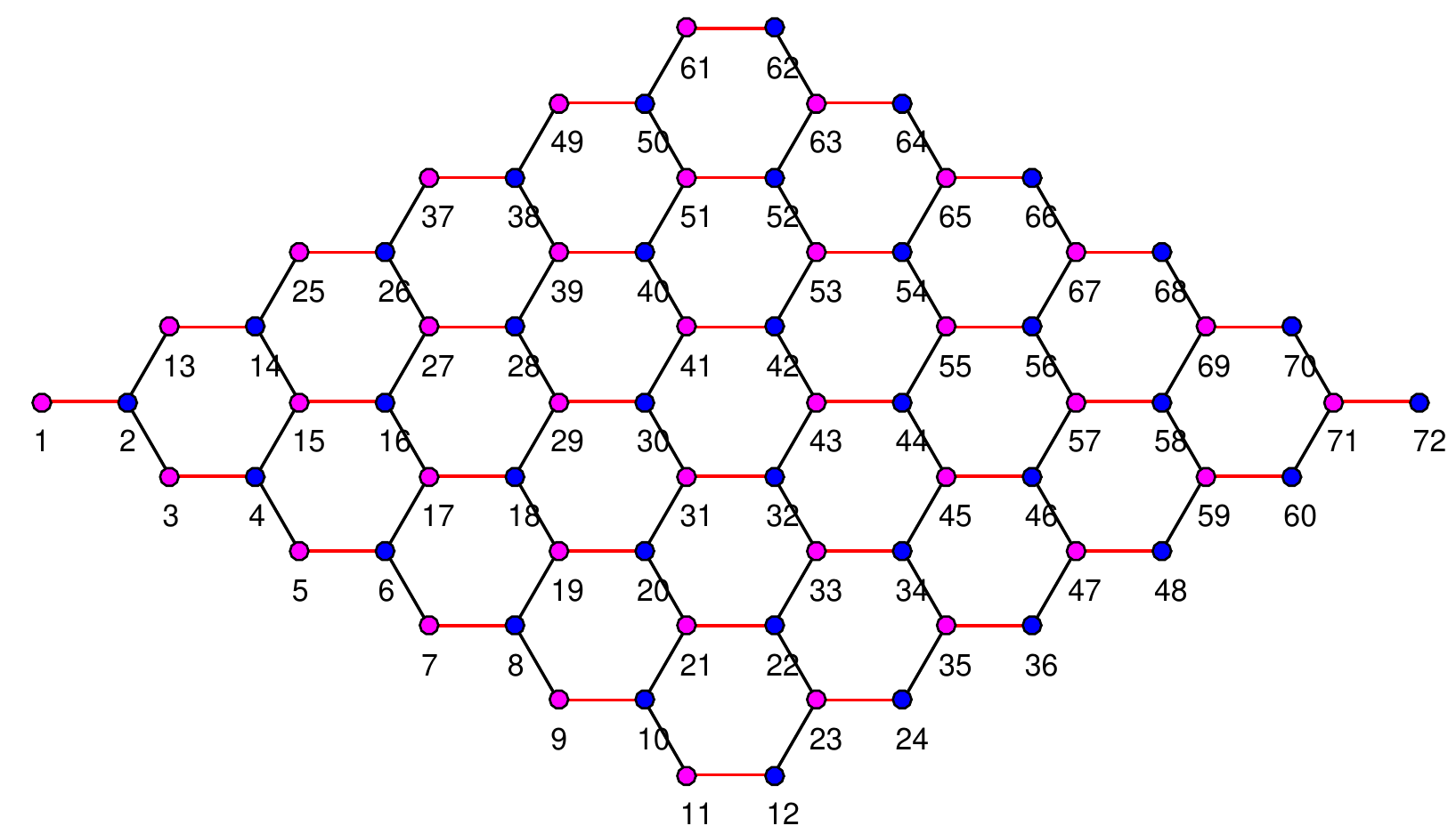}
\caption{The geometry of
the $L=6$ strained honeycomb lattice with 72 sites, where red and blue label A and B sublattice, respectively. Stress is added along the x-direction as shown by red lines, which $t_2=t-\Delta{t}$. The dark lines indicate $t_{1,3}=t$. Here $t$ is the nearest hopping integrating term and $\Delta{t}$ represents the strength of strain.}
\label{Fig:structure}}
\end{figure}

Graphene is a 2D sheet of carbon atoms packed hexagonal structure as depicted in Figure.~\ref{Fig:structure}. Since its discovery\cite{Novoselov2004}, graphene has stimulated a tremendous burst of research activities on fundamental and applied grounds. Graphene has multiple astonishing performance in several areas, which is the toughest 2D material ever measured \cite{Lee385}, and also has extremely high charge carriers mobility (charge carriers move at a speed of $c/300$ ($c$ is the speed of light in vacuum)\cite{Novoselov2005Two,PhysRevLett.100.016602} and behave like massless particles) More than these, graphene is
impermeable to standard gases\cite{Bunch2008Impermeable} and it could be a perfect thermal conductor\cite{Balandin2008Superior}. Numerous other honeycomb liked systems have attracted a lot of attention, such as silicene\cite{PhysRevLett.108.155501}, germanene\cite{PhysRevLett.102.236804} and phosphorene\cite{Li2014Black}, which also have excellent properties like graphene.
Thus, we expect that these honeycomb like materials will pave the way toward a new age of material.

There is no gap in the band structure of graphene which exhibits semi-metallic properties\cite{NOVOSELOV20081,Peres2010The}. However, whether the energy gap could be modified in real graphene material to realize the metal-insulator transition (MIT)\cite{PhysRevLett.102.026802,PhysRevLett.111.036601,PhysRevLett.111.056801} is still controversial. If it is possible to control the opening and closing of the energy gap in real graphene, the application of graphene in the electronic device will be greatly promoted.

Recent theoretical and experimental work shows that hydrogenation\cite{Elias610}, application of stress\cite{doi:10.1021/nn800459e,PhysRevB.81.081407,PhysRevLett.102.056808,RevModPhys.88.025005,PhysRevB.80.045401}, forming nanoribbons\cite{PhysRevLett.98.206805}, Moir$\acute{e}$-stripe modulation\cite{Hunt1427,Ponomarenko2011Tunable}, etc. can be used to open the band gap and achieve MIT in graphene. In this work, we explore different approaches to achieving insulating state, considering the interactions between electrons. Graphene is clearly a semimetal without any electron-electron interactions, but it will turn into an insulating antiferromagnet(AFM) under very strong interactions\cite{RevModPhys.84.1067}. However, it still remains unclear as to what we should expect in real graphene materials. For instance, in the real case, it is common to find stress effect in spite of interaction effects, and it is an amazing thing that the electronic properties could be modified by mechanical forces\cite{Gui2015Local,Aslani2015Characterization,Fanbanrai2015Effects,
PhysRevB.96.085126,Choi2009Effects,Cosma2014Strain,Craco2015Revealing}. As for graphene, defects could be introduced naturally or intentionally while growing. The existence of these defects also attract wide interest\cite{Summerfield2016Erratum,Zhang2016Lattice}. In the experiment, strain could be measured and detected by atomic force microscopy. And strain of $25\%$ on the graphene can be achieved\cite{Lee385}, so that the system could completely come into the antiferromagnetic region\cite{Lee385}.
Theoretically, there are also a lot of work based on the tight-binding model, studying the effect of uniaxial strain on the properties of graphene\cite{PhysRevB.80.045401}. It seems that the gap suddenly arises when deformation is beyond $20\%$ without a transition region\cite{PhysRevB.80.045401,PhysRevB.81.081407}. To illustrate the strain effect in these heavy fermion materials, the electron interactions should also be considered, and it is interesting to find that the quantum critical point (QCP) strain value is suppressed by interactions\cite{PhysRevLett.115.186602}. The localization effect and mutual effect of these two factors are the key to understanding the mechanism of MIT in strained graphene. In this paper, we will focus on the competition between electronic correlations and uniaxial strains, and the phase diagram is also discussed.

% --------------------------------------------------------------------

\noindent

\section{Model and numerical method}
The Hubbard model\cite{Hubbard} is believed to be a versatile paradigm to study strongly correlated electrons on a lattice. It is effective to describe electrons in partially filled energy bands. The parameters of the model, such as hopping and electron-electron interaction, can be considered to be generated by integrating the corresponding degrees of freedom of all energy bands except partially filled narrow bands. Thus, this model\cite{RevModPhys.70.1039} can be used to explain the interaction-driven transition between conducting and insulating behaviors.

To shed light on the physical problems involved in the strained-correlation fermions, we use determinant quantum Monte Carlo (DQMC) to study the half-filled repulsive Hubbard model on the honeycomb lattice.
In the Dirac electronic correlation system, it is already known that there exist a QCP for MIT at around $Uc\sim3.89t$\cite{PhysRevB.72.085123,Sorella2012Absence,
PhysRevX.3.031010,PhysRevB.92.045111}. In this strength of interaction region, our numerical DQMC method can conduct an accurate simulation.
We consider the strained Hubbard Hamiltonian on the honeycomb lattice as
\begin{eqnarray}
H =&&-\sum_{\bfi\eta\sigma}
       t^{\phantom{\dagger}}_{\eta}
       a^{\dagger}_{\bfi\sigma}b^{\phantom{\dagger}}_{\bfi+\eta\sigma} +h.c.
-\mu\sum_{\bfi\sigma}\left(n_{\bfa\bfi\sigma}+n_{\bfb\bfi\sigma}\right) \nonumber\\
     &&+ U\sum_\bfi \left( n_{\bfa\bfi\uparrow}n_{\bf{ai}\downarrow}+ n_{\bf{bi}\uparrow}n_{\bf{bi}\downarrow}\right).
   \label{eq:model}
\end{eqnarray}

Here $a^\dag_{\bfi\sigma}$ ($a_{\bfi\sigma}$) are the spin-$\sigma$
electron creation (annihilation) operators at site $\bfi$ on sublattice A. And $b^\dag_{\bfi\sigma}$ ($b_{\bfi\sigma}$) are operators acting on sublattice B. $U > 0$ is the
on-site Coulomb repulsion. $t_{\eta}$ denotes the hopping integral
between two nearest-neighbor sites.
The chemical potential $\mu$
determines the average density of the system.
$n_{\bf{ai}\sigma}=a^{\dagger}_{\bfi\sigma}a^{\phantom{\dagger}}_{\bfi\sigma}$ and
$n_{\bf{bi}\sigma}=b^{\dagger}_{\bfi\sigma}b^{\phantom{\dagger}}_{\bfi\sigma}$
are the number operators.  Strain is introduced through the hopping matrix elements
$t_{\eta}$ along the x-direction which is shown by red lines in Fig.~\ref{Fig:structure}. $t_{1,3}=t$ and $t_2=t-\Delta{t}$. Here $\Delta{t}$ represents a
measure of stress strength. We take $t=1$ as the scale of energy in our system.
In this work, we set $\mu=0$, thus, the system is half-filled by electrons. And in this case, the Hamiltonian will be particle-hole symmetric even with strain on it, so that the Hamiltonian is sign-problem free to solve in our DQMC simulations.

In the finite-temperature DQMC method, it follows the strategy that take the partition function as an integral over all possible configurations which is carried out by random Monte Carlo sampling. To be more specifically, we can get static and dynamic observables at an given temperature $T$ through this approach.  As the system is sign-problem free due to the particle-hole symmetry, our calculations can still be in high numerical precision at large enough $\beta=1/T$, thus, we can converge the data to get the ground state. In this work, 4000 sweeps were used to get to the equilibrium state, and 12000 additional steps were made to measure the system. The periodic boundary conditions were used in the simulation, and all the results are obtained on the $2\times 12^2$ honeycomb lattice with is large enough to consider the size effect. Fig.~\ref{Fig:structure} shows the $L=6$ geometry with strain on it.
% --------------------------------------------------------------------

\begin{figure}
\center{%\includegraphics[width=\linewidth]{Fig1_structure}
\includegraphics[width=3.0in]{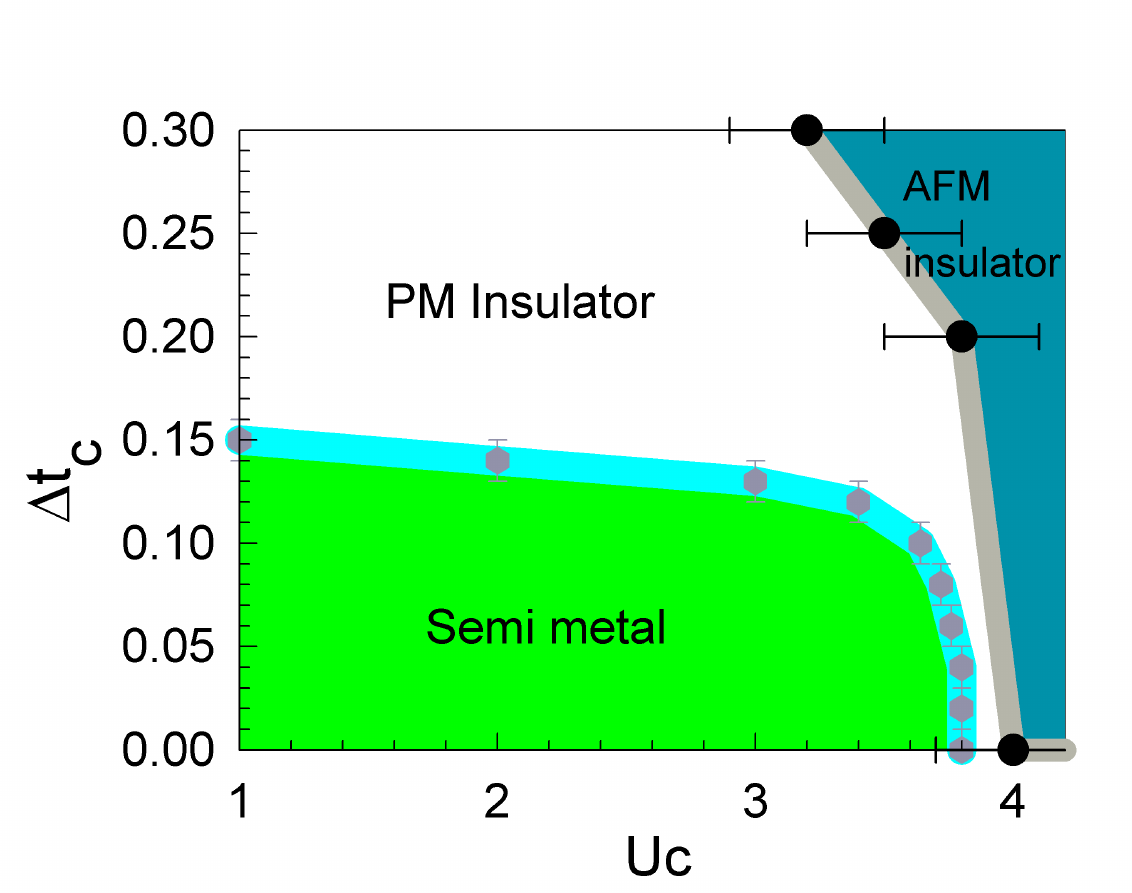}
\caption{Phase diagram of the strained Hubbard model on the honeycomb lattice at half-filling. $\Delta t$ labels the strain strength and $U$ represents the local Coulomb repulsion.
The metallic phase boundary is determined by the temperature dependence
of the conductivity $\sigma_{dc}$. The metallic region is filled with green color and the (grey) hexagonal mark point is obtained from our DQMC simulation results. The blue region on the left indicates the existence of long-range anti-ferromagnetic order.}
\label{Fig:PhaseDiagram}}
\end{figure}

We mainly discussed the MIT in this article. The temperature-dependent DC conductivity $\sigma_{dc}(T)$  is an intuitive physical observation to help characterize the MIT. So that we focus on discussing $\sigma_{dc}(T)$ behavior in our case. According to the fluctuation-dissipation theorem, in the zero frequency limit, $\sigma_{dc}(T)$ is related to the current-current correlation function. As we could do imaginary-time simulations using DQMC, the real-frequency quantities can be easily obtained through analytic continuation methods. In our case, we adopted an approximation \cite{Trivedi1995} to define $\sigma_{dc}(T)$, which has been widely used as a benchmark in previous work \cite{Trivsig,Trivedi1995,Denteneer1999}.

The definition of DC conductivity is:
\begin{equation}
 \sigma_{dc}(T) =
   \frac{\beta^2}{\pi} \Lambda_{xx} ({\bf q}=0,\tau=\beta/2).
 \label{eq:condform}
\end{equation}
In Equation (\ref{eq:condform}), $\Lambda_{xx}$ is the current-current correlation function, and it can be expressed as $\Lambda_{xx} ({\bf q},\tau) = \ob{ j_x ({\bf q},\tau) \, j_x
(-{\bf q}, 0)}$, where $j_x ({\bf q},\tau)$ is the Fourier transform of $j_x(\bfr,\tau)$ (time-dependent current operator) along the $x$-direction. When we focus on studying the system in the temperature lower than the energy scale, in which the density of states has a significant structure, the DC conductivity definition, in Equation(\ref{eq:condform}), can be very practicable considering such approximation. The applicability has been checked in our recent work of the Hubbard model on the honeycomb lattice\cite{PhysRevLett.120.116601}. We can get the MIT critical strength value $U_{c}$\cite{PhysRevLett.83.4610} within the formula from the change of low-$T$ behavior of $\sigma_{dc}(T)$.
We also investigate the magnetic properties by carrying out measurements of spin-spin correlation functions and the spin structure factor,
\begin{equation}
S_{AF}=\frac{1}{N_c}\langle( \sum_{\bf r\in A} \hat{S_{\bf r}^{z}}-\sum_{\bf r\in B}\hat{S_{\bf r}^{z}} )^2\rangle.
\label{eq:Saf}
\end{equation}
where $N_c$ is the number of sites in the lattice, and $A$ and $B$ represents the sublattices of the honeycomb lattice. Here $\hat{S_{\bf r}^{z}}$ is the $z$ component spin operator. The results are normalized by $\langle ... \rangle$ including all the lattice sites. The antiferromagnetic results present in this work are concluded from the constrained-path quantum Monte Carlo (CPQMC) method\cite{PhysRevLett.78.4486,PhysRevLett.74.3652,PhysRevB.84.121410}, which uses the constrained-path approximation in the simulation procedure without a sign problem.

\begin{figure}
\center{%\includegraphics[width=\linewidth]{Fig1_structure}
\includegraphics[width=3.2in]{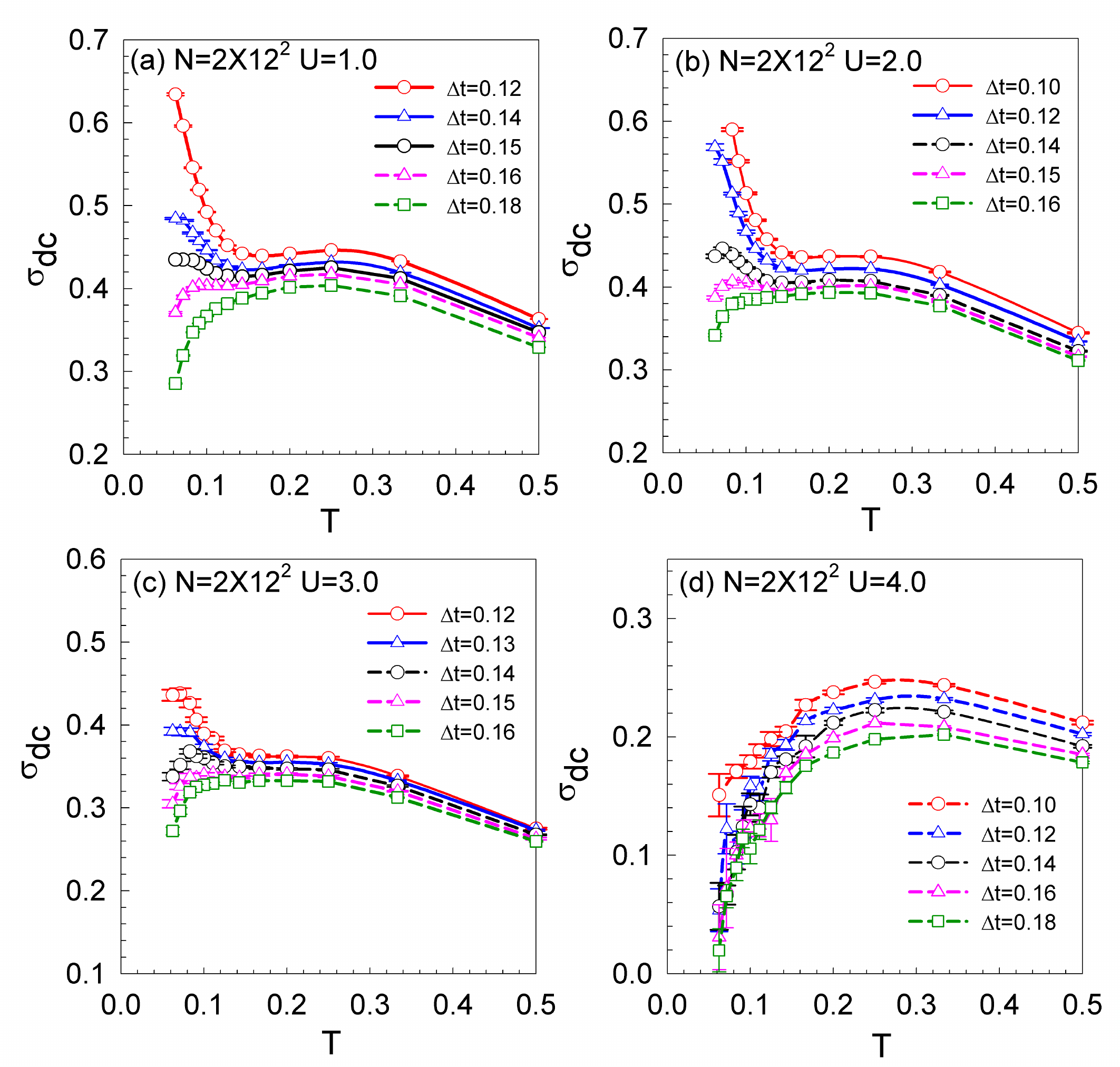}
\caption{Temperature dependence of the DC conductivity $\sigma_{dc}$
measured on the $L=12$ lattice with strain along the x-direction. $N=2\times L^2$ is the total site numbers. Panels correspond to
different couplings: (a) $U=1.0$, (b) $U=2.0$, (c) $U=3.0$ and (d) $U=4.0$. In each figure, lines are guides to the eyes. $\Delta{t}$ represents for the strain strength. Metallic and
insulating behaviors are indicated by solid and dashed lines
respectively. The MIT behavior can be easily observed from the low-$T$ behavior of $\sigma_{dc}$, whereas in panel (d) the insulating behavior is preserved by strain effect.
}
\label{Fig:strainU1_U4}}
\end{figure}
\section{Results and discussion}
Our simulations are mainly computed on the $L = 12$ lattice considering the size effect. $N=2\times L^2$ is the total site numbers. The quantity of immediate interest in this work is to study the possible MIT of the strained honeycomb lattice, which can be measured by the $conductivity$ and its $T$ dependence behavior. So we show the $\sigma_{dc}(T)$ behavior at low temperature here in Figure.~\ref{Fig:strainU1_U4}. For some curves, $d\sigma_{dc}(T)/dT<0$ and $\sigma_{dc}(T)$ diverges as $T$ decreases to the $T\rightarrow0$ limit, which suggesting metallic behavior of the system. In contrast, the system is insulating when $d\sigma_{dc}(T)/dT>0$ at $T\rightarrow0$ limit. Thus, we can easily tell the MIT point of the strained system.
In panels (a)-(b) of Fig.~\ref{Fig:strainU1_U4}, we described $\sigma_{dc}(T)$ behaviors at low coupling strengths under several strain strengths. To understand the role of strain on the conductivity, we can take a careful analyzation of panel (a) in Fig.~\ref{Fig:strainU1_U4}. In spite of $\Delta{t}$ strength, the conductivity increases as temperature increases until $T\succeq0.25$. However when $U = 1.0$ and with a strong enough strain strength $\Delta{t} = 0.20$, the curve bent down when temperature drops and $\sigma_{dc}(T)$ approaches zero as $T\rightarrow0$, suggesting insulating behavior of the system. In contrast, when we take a look at $\Delta{t} = 0.10$, the red line diverges with feature of $d\sigma_{dc}(T)/dT<0$ for $T\leq0.2$, characteristic the metallic behavior of the system. Thus, we can clearly get the strain-driven MIT critical point from the panels. At $U=1.0$, the critical $\Delta{t}$ value can be read as $\Delta{t_{c}}\sim 0.15\pm 0.01$. As for panel (b) in Fig.~\ref{Fig:strainU1_U4} with $U=2.0$. At $\Delta{t}=0.12$, $\sigma_{dc}(T)$ curve is concave and $d\sigma_{dc}(T)/dT<0$ for $T\leq0.2$. When $T$ drops to around $0.1$, $\sigma_{dc}(T)$ increases rapidly, in other words, the system is metallic. At $\Delta{t}=0.14$, in contrast, the conductivity decreases as the temperature decreases, which indicating insulating behaviors. Thus, there exist a MIT critical strain strength at around $\Delta{t}=0.14\pm0.01$ for $U=2.0$ on the graphene lattice.
Then it moves to the larger interaction case, as shoen in Fig.~\ref{Fig:strainU1_U4}(c) and (d). When $U=3.0$, the system is still metallic without considering strain effect. So that under small strain strength, it is still possible that the graphene can stay in metallic region. At low $T$ for $T<0.1$, the sign of $d\sigma_{dc}(T)/dT$ is negative when $\Delta{t}\leq0.12$, but it switches to positive when $\Delta{t}$ is larger than $0.13$. Therefore, the MIT critical  $\Delta{t}$ strengths can be read as $\Delta{t_{c}}\sim 0.13\pm 0.01$ for $U=3.0$. As for $U=4.0$ which is larger than the critical $U_{c}=3.89$ for MIT in the graphene lattice without strain, the system is an insulator. From panel (d), we can see all the $\sigma_{dc}(T)$ curves are bent down when temperature decreased for $T\rightarrow0$. The value of $\sigma_{dc}(T)$ gets suppressed when strength of strain increases, that is to say the insulating behavior is preserved by the strain effect in this case. To describe the metal-insulator phase boundary more precisely, we closely studied the conductivity behaviour around $U\sim3.8$. Thus, according to Fig.~\ref{Fig:strain_dc_U} in Appendix, we make the metal-insulator phase boundary more smooth, as shown in the phase diagram Fig.~\ref{Fig:PhaseDiagram}.

\begin{figure}
\center{%\includegraphics[width=\linewidth]{Fig1_structure}
\includegraphics[width=3.2in]{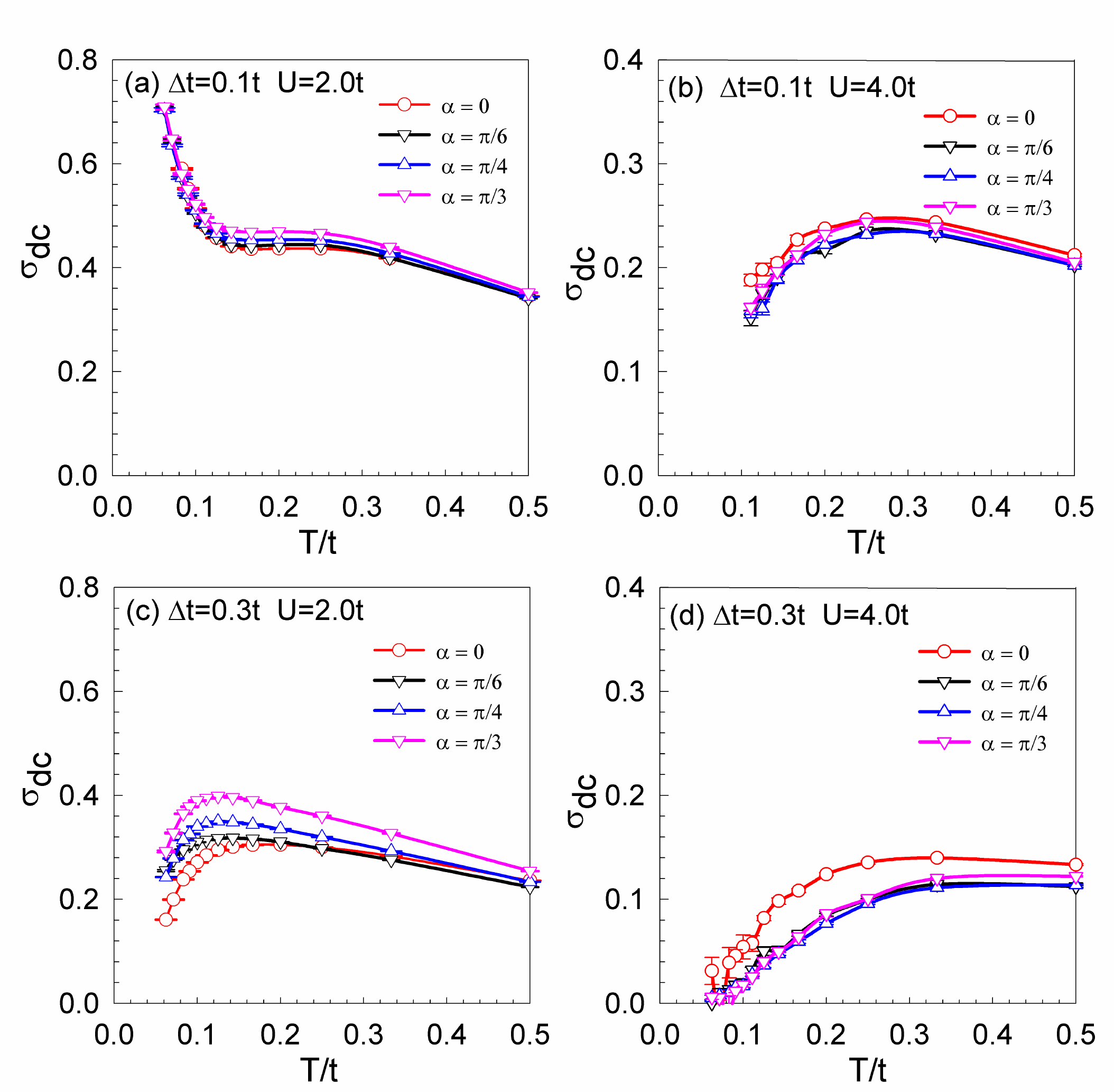}
\caption{
Temperature dependence of the DC conductivity $\sigma_{dc}$
measured on the $L=12$ lattice with strain along different directions. Panels correspond to different couplings and strain strengths: (a) $U=2.0$, $\Delta{t}=0.1$ is in the metallic region, and (b) $U=4.0$, $\Delta{t}=0.1$, (c) $U=2.0$, $\Delta{t}=0.3$, (c) $U=4.0$, $\Delta{t}=0.3$ are in the insulating region. In each figure, lines are guides to the eyes. $\Delta{t}$ represents for the strain strength, $\alpha$ is the inclined angle between strain and $x$-axis.
}
\label{Fig:strain_angle}}
\end{figure}

%\begin{figure}
%\includegraphics[scale=0.41]{strainU3_U4.pdf}
%\caption{Temperature dependence of the DC conductivity $\sigma_{dc}$
%measured on the $L=12$ lattice with strain along x-direction. Panels correspond to
%different couplings: (a) $U=3.0$, (b) $U=4.0$. In each figure, lines are guides to the eyes. $\Delta{t}$ represents for the strain strength. Metallic and
%insulating behaviors are indicated by solid and dashed lines
%respectively. And the low-$T$ behavior of $\sigma_{dc}$ in panel (a) clearly indicates the strain-driven metal-insulator transition, while in panel (b) the insulating behavior is preserved by strain effect.
%}
%\label{Fig:strainU3_U4}
%\end{figure}

In Fig.~\ref{Fig:strain_angle}, we also investigate the effect of the orientation of the strain. Here $\alpha$ in each panel is the angle with x-axis. Panel (a) is in the metallic regime for $U=2.0$ and $\Delta{t}=0.1$. As $\Delta{t}$ increases to $0.3$, the system turns into the insulating regime, as panel (c) shows. Comparing panel (b) and (d), the result at $U=4.0$ is shown that graphene stays in the insulating state no matter how strong the strain applied. According to the lines in Fig.~\ref{Fig:strain_angle}, it seems that the behavior of $\sigma_{dc}$ is pretty similar when $\alpha$ changes. We only consider the nearest hopping in our Hamiltonian, which is a single-shell model. The shell behavior to the angles is very similar to the first shell of the model including up to third-nearest-neighbor interactions\cite{doi:10.1021/acs.jpcc.8b04502}. The angular dependence is not very important considering atoms in the second and third shells, thus, we get the similar behaviors of $\sigma_{dc}(T)$ here with different orientations of strain. Therefore, we can focus our study on the strength of strain and ignore its orientation when exploring the strain effect on MIT.

\begin{figure}
\center{%\includegraphics[width=\linewidth]{Fig1_structure}
\includegraphics[width=3.2in]{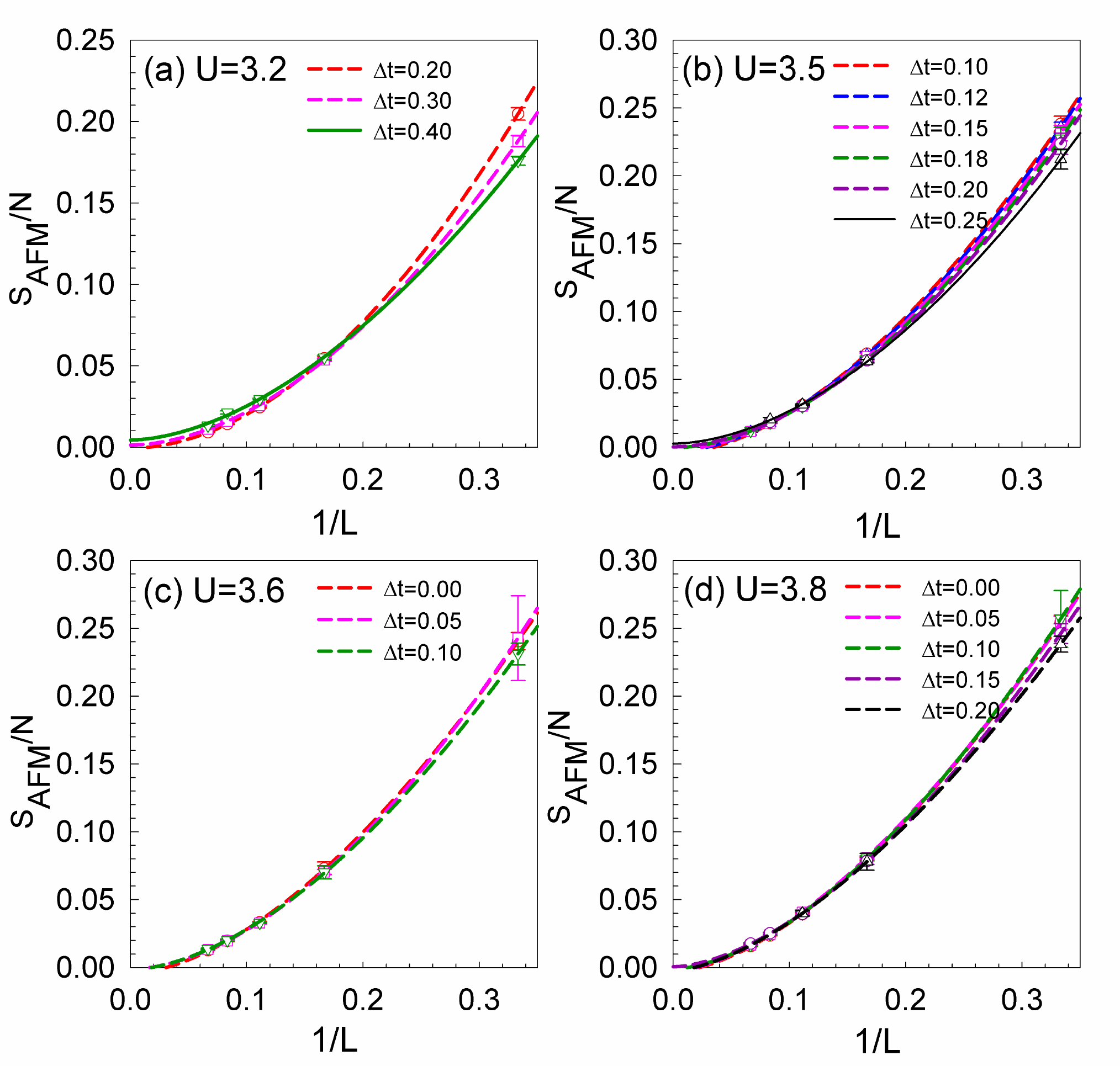}
\caption{
Finite-size scaling results of the AF spin structure factor by CPQMC. $S_{AF}$ is plotted as a function of $1/L$ corresponding to various value of interaction strength. A finite $S_{AF}$ in the $L\rightarrow\infty$ limit indicates the existence of long-range magnetic order. Symbols on the lines are simulation results on $L=3,6,9,12,15$ lattice sizes within statistical errors. The lines are cubic polynomial fits to the data.
}
\label{Fig:strain_AFM_U34}}
\end{figure}

Further more, we also studied the effect of strain on long-range magnetic order. Data shown in  Fig.~\ref{Fig:strain_AFM_U34} and Fig.~\ref{Fig:strain_AFM_U2345} are AF spin structure factor on lattices up to $L=15$. For small $U=2,3$, there is no long-range order when $U<U_{c}\sim3.89t$\cite{PhysRevB.72.085123,Sorella2012Absence,
PhysRevX.3.031010,PhysRevB.92.045111}. Even when $\Delta{t}=0.30t$, there is no AF order according to Fig.~\ref{Fig:strain_AFM_U2345}(a) and (b). As for small lattices, the correlation strength of antiferromagnetism decreases while strain increases, which indicates that in the weak interaction case, $S_{AF}$ are suppressed by strain. In contrast, in Fig.~\ref{Fig:strain_AFM_U2345}(c) and (d), AF correlation strength is enhanced by strain at large $U$. The AF order is protected by strain at strong interaction ($U>U_{c}$). As Fig.~\ref{Fig:strain_AFM_U34} shows, it is interesting to note that the AF order could exist even when $U<U_{c}$. Thus, there is a transition point that exists for long-range order by strain effect. We obtained (a)$U=3.2, \Delta{t}=0.30$, (b)$U=3.5, \Delta{t}=0.25$,(a)$U=3.6, \Delta{t}=0.25$ and (a)$U=3.8, \Delta{t}=0.20$ from the finite size scaling results. Based on the analysis earlier, the paramagnetic-antiferromagnetic transition boundary could be summarised for nonzero $\Delta{t}$, as shown in Fig.~\ref{Fig:PhaseDiagram}.

\section{Summary}
We have carefully studied the electronic and magnetic properties of the strained Hubbard model on the honeycomb lattice. Using DQMC, we characterize the $conductivity$ with the $\sigma_{dc}(T)$ behavior at low temperature to explore the mechanism of strain-driven MIT in the honeycomb lattice. According to our simulations, it is clear to see that the conductivity $\sigma_{dc}$ is suppressed by strain effect. When strain is applied, the value of hopping parameter along the strain direction decreases from $t$ to $t-\Delta{t}$, thus, the critical electron-electron interaction strength $U$ for MIT reduces. And the critical strain strength decreases as the local Coulomb repulsion $U$ enhanced. In other words, under the interaction of strain and Coulomb correlation, the electrons get localized more effectively, which expands the insulating phase and gives us a strain-interaction driven MIT phase diagram.

As for the magnetic properties, we found that strain could also enhance the AF order at around $U_{c}$ in the clean limit. While strain is affected, we observed the appearance of the antiferromagnetic order phase. It seems that strain can protect the long antiferromagnetic order when $U$ is strong. In conclusion, we show that strain-interaction-driven graphene could be a promising rout to achieve MIT in graphene, and the phase diagram reported in this work could be used as a guidance in the modulation of conductivity in real graphene materials.
% Experimental section

%\section{Experimental Section}
%\threesubsection{First part of experimental section}\\
%\threesubsection{Second part of experimental section}\\

%\medskip
%\textbf{Supporting Information} \par %Please delete the Suppporting Information statement if it is not applicable. Please supply Supporting Information in another file. Supporting information should not be provided in .tex format
%Supporting Information is available from the Wiley Online Library or from the author.

% Acknowledgements
%\medskip
%\textbf{Acknowledgements} \par %delete if not applicable))
T. M. and L. F. Z were
supported by NSFCs (No. 11774033 and 11974049) and Beijing Natural Science Foundation (No. 1192011).% We acknowledge computational support from the
%Beijing Computational Science Research Center (CSRC), the support of
%HSCC of Beijing Normal University.
%\end{acknowledgments}

\bibliography{reference}

\appendix

\begin{figure}
\center{%\includegraphics[width=\linewidth]{Fig1_structure}
\includegraphics[width=3.2in]{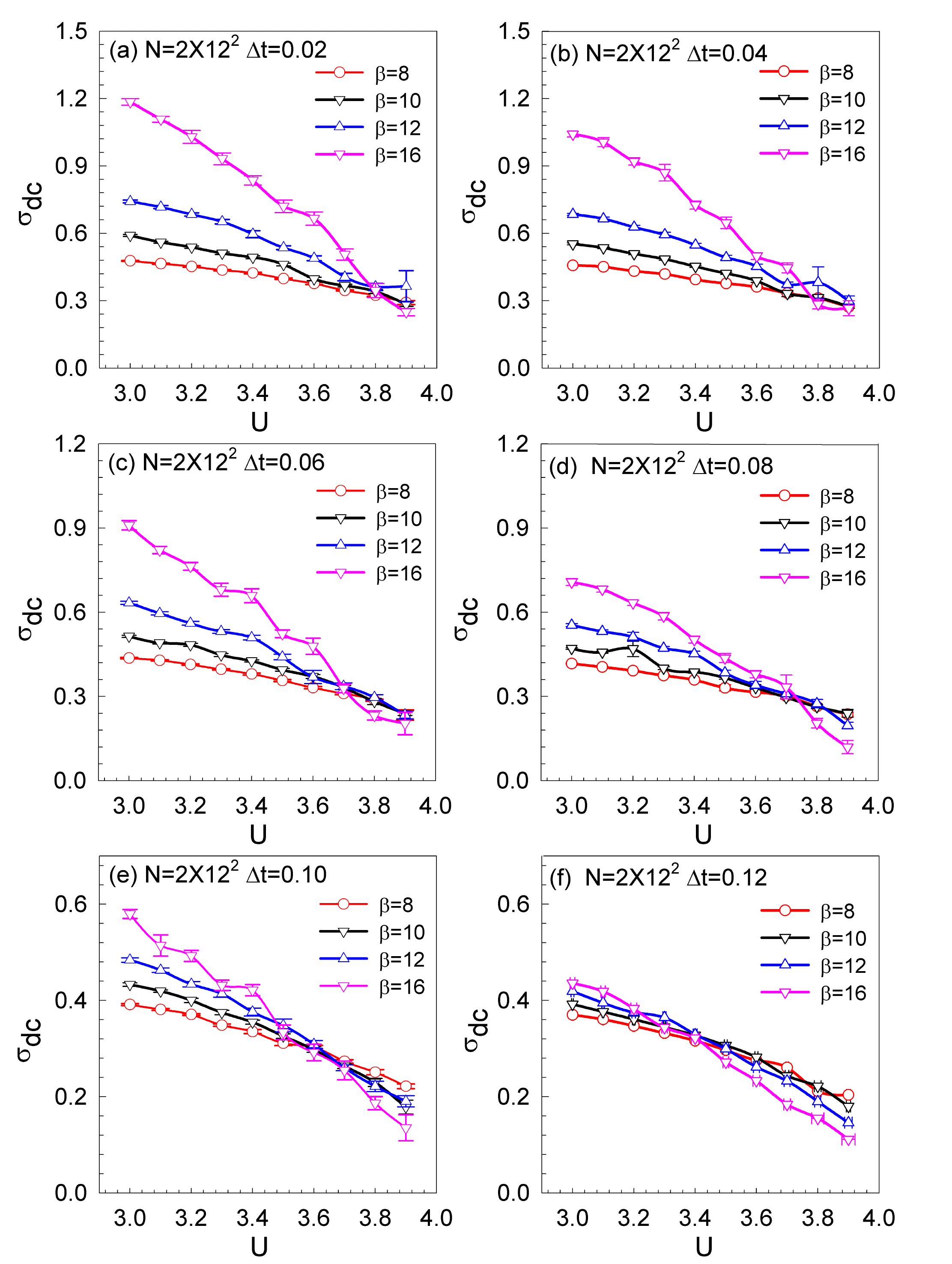}
\caption{$\sigma_{dc}$ behavior as the interaction changes at different temperatures. Panels correspond to
different strain strengths from $\Delta{t}=0.02$ to $\Delta{t}=0.12$. The crosspoint in each panel indicates the transition point: (a) $\Delta{t}=0.02$,$U\simeq3.80$, (b) $\Delta{t}=0.04$,$U\simeq3.80$,(c) $\Delta{t}=0.06$,$U\simeq3.75$,(d) $\Delta{t}=0.08$,$U\simeq3.70$,(e) $\Delta{t}=0.10$,$U\simeq3.65$ and (f) $\Delta{t}=0.12$,$U\simeq3.40$.
}
\label{Fig:strain_dc_U}}
\end{figure}

\subsection{Computing the DC conductivity}
 In this work, the low-temperature behavior of DC conductivity $\sigma_{dc}$ is used to distinguish metallic or insulating phases. We found that the transition point changes dramatically at around $U\sim 3.80$. So in appendix Fig.~\ref{Fig:strain_dc_U}, we carefully checked the phase boundary by plotting the $\sigma_{dc}(U)$ curves, which indicate a critical transition strength $U_{c}$ in each panel. While in panel (a), when $U<3.80$, $\sigma_{dc}$ gradually increases as $\beta$ get larger, which indicates the metallic behavior. On the other side, when $U>3.80$, $\sigma_{dc}$ decreases as $\beta$ increases, which act as an insulator. Thus, the critical transition point for $\Delta{t}=0.02$ is $U\simeq3.80$. And other transition points are (b) $\Delta{t}=0.04$,$U\simeq3.80$,(c) $\Delta{t}=0.06$,$U\simeq3.75$,(d) $\Delta{t}=0.08$,$U\simeq3.70$,(e) $\Delta{t}=0.10$,$U\simeq3.65$ and (f) $\Delta{t}=0.12$,$U\simeq3.40$.

\begin{figure}
\center{%\includegraphics[width=\linewidth]{Fig1_structure}
\includegraphics[width=3.2in]{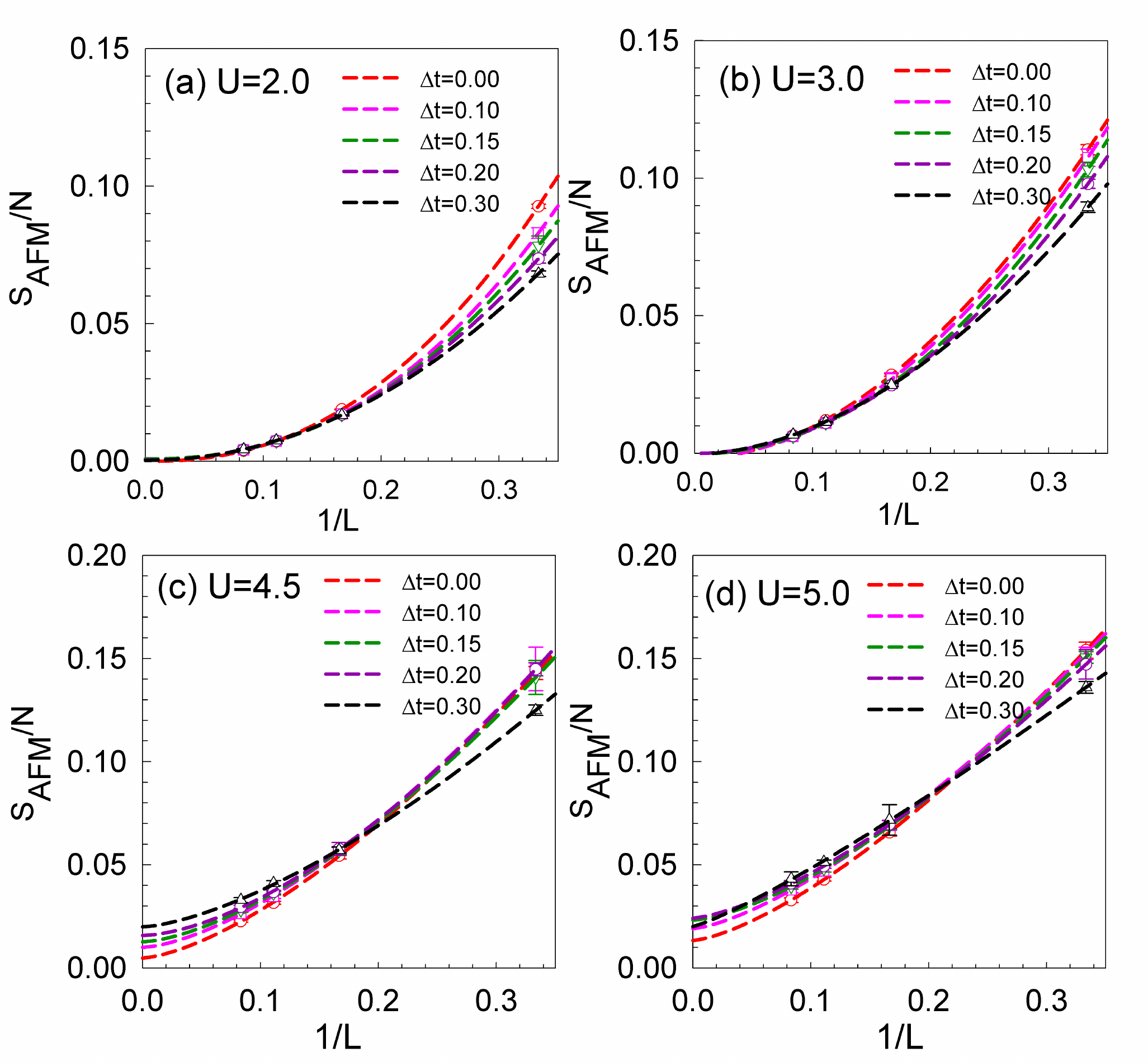}
\caption{
Study of the existence of long-range magnetic order in strained system at various $U$ values.
}
\label{Fig:strain_AFM_U2345}}
\end{figure}

\subsection{Existence of the AF order phase}
To establish the phase diagram, the finite-size effect on the AF spin structure factor $S_{AF}$ has been carefully examined in the manuscript. We extrapolated the data to the thermodynamic limit to get the order parameters. As shown in Fig .~\ref{Fig:strain_AFM_U2345}, we discussed the strain effect on the honeycomb lattice with various interaction strengths. All the results are summarized in the phase diagram Fig .~\ref{Fig:PhaseDiagram}.

\end{document}